\documentclass[aps,prl,twocolumn,10pt,amsmath,amssymb,bibnotes,superscriptaddress,longbibliography
]{revtex4-1}

\usepackage{graphicx, braket}
\usepackage{siunitx}
\usepackage{xcolor}
\usepackage{pdfpages}
\usepackage[colorlinks,citecolor=blue,linkcolor=blue]{hyperref}

\newcommand{\mb}{moir\'e band }
\newcommand{\m}{moir\'e }
\newcommand{\bperp}{\ensuremath{B_\perp}}
\newcommand{\bpar}{\ensuremath{B_\parallel}}
\newcommand{\delnu}{\ensuremath{\nu_H-\nu}}
\newcommand{\idc}{\ensuremath{I_{dc}}}
\newcommand{\ctwo}{\ensuremath{\mathcal{C}_2}~}

\newcommand{\rxx}{\ensuremath{R_{xx}}}

\usepackage{xcolor}

\usepackage[normalem]{ulem}

\makeatletter
\AtBeginDocument{\let\LS@rot\@undefined}
\makeatother

\begin{document}

\title{Superconductivity in Twisted Double Bilayer Graphene Stabilized by WSe$_2$}
 \author{Ruiheng Su}
  \affiliation{Stewart Blusson Quantum Matter Institute, University of British Columbia, Vancouver, British Columbia, V6T1Z4, Canada}
  \affiliation{Department of Physics and Astronomy, University of British Columbia, Vancouver, British Columbia, V6T1Z1, Canada}
\author{Manabendra Kuiri}
   \affiliation{Stewart Blusson Quantum Matter Institute, University of British Columbia, Vancouver, British Columbia, V6T1Z4, Canada}
   \affiliation{Department of Physics and Astronomy, University of British Columbia, Vancouver, British Columbia, V6T1Z1, Canada}
 \author{Kenji Watanabe}
   \affiliation{Research Center for Functional Materials, National Institute for Materials Science, Namiki 1-1, Tsukuba, Ibaraki 305-0044, Japan}
 \author{Takashi Taniguchi}
   \affiliation{International Center for Materials Nanoarchitectonics,
  National Institute for Materials Science, Namiki 1-1, Tsukuba, Ibaraki 305-0044, Japan}
 \author{Joshua Folk}
 \email{jfolk@physics.ubc.ca}
	\affiliation{Stewart Blusson Quantum Matter Institute, University of British Columbia, Vancouver, British Columbia, V6T1Z4, Canada}
	\affiliation{Department of Physics and Astronomy, University of British Columbia, Vancouver, British Columbia, V6T1Z1, Canada}
\date{\today}

\begin{abstract}
Superconductivity has been previously observed in magic-angle twisted stacks of monolayer graphene but conspicuously not in twisted stacks of bilayer graphene, although both systems host topological flat bands and symmetry-broken states. Here, we report the discovery of superconductivity in twisted double bilayer graphene (TDBG) in proximity to WSe$_2$.  Samples with twist angles 1.24$^\circ$ and 1.37$^\circ$ superconduct in small pockets of the gate-tuned phase diagram within the valence and conduction band, respectively. Superconductivity emerges from unpolarized states near van Hove singularities and next to regions with broken isospin symmetry, showing the correlation between a high density of states and the emergence of superconductivity in TDBG while revealing a possible role for isospin fluctuations in the pairing.

\end{abstract}
\maketitle
Identifying the essential components of superconductivity in graphene-based systems remains a critical problem in 2D-materials research, connecting this field to the mysteries that underpin investigations of unconventional superconductivity throughout condensed-matter physics.  
Superconductivity in graphene heterostructures
is consistently correlated with flat electronic bands, whether induced by a \m potential, as in stacks of  graphene monolayers with alternating twist\cite{cao2018unconventional, park2021tunable,hao2021electric, park2022robust,zhang2022promotion} such as magic-angle twisted bilayer graphene (TBG), or by gate voltage as in rhombohedral trilayer graphene (RTG) \cite{zhou2021superconductivity} or Bernal bilayer graphene (BBG)\cite{zhou2022isospin} . Beyond the connection between flat bands and superconductivity\cite{bistritzer2011moire,balents2020superconductivity}, little is agreed upon in terms of the pairing mechanism or symmetry in each system, or about the characteristics of the band structure that are crucial to the emergence of superconductivity\cite{you2022kohn,chou2021correlation, lian2019twisted, wu2018theory, khalaf2021charged, you2019superconductivity,lee2019theory, balents2020superconductivity}. Introducing a sheet of tungsten diselenide (WSe$_{2}$) next to  graphene has been reported to stabilize superconductivity in both TBG and BBG\cite{arora2020superconductivity,zhang2022spin}, enhancing critical temperatures and fields and extending the parameter range over which superconductivity emerges, but the mechanism behind this effect remains a subject of debate.

There is a puzzling contrast between the consistent observation of superconductivity in TBG\cite{cao2018unconventional,lu2019superconductors} and none so far in twisted double bilayer graphene (TDBG, two twisted Bernal bilayers)\cite{shen2020correlated,liu2020tunable, caoTunableCorrelatedStates2020,he2021symmetry,burg2019correlated}. The two systems have similar \m band widths\cite{PhysRevB.100.201402}, similar signatures of strong interactions including correlated insulators and broken-symmetry states\cite{cao2018correlated,shen2020correlated,sharpe2019emergent,liu2020tunable,burg2019correlated, caoTunableCorrelatedStates2020, he2021symmetry, kuiri2022}, and both have \m bands that are understood to be topological, with Chern number $|C|=1$ for the first \mb of TBG and $|C|=2$ for TDBG\cite{PhysRevB.99.235406,PhysRevB.99.075127}. One difference is the lack of \ctwo symmetry (two-fold rotation) in TDBG, whereas \ctwo symmetry is preserved in the family of alternating-twist graphene monolayer stacks.

 The high degree of tunability in TDBG would make it a powerful experimental probe of the mechanisms underlying graphene superconductivity, with top and back gates to control the band structure through the vertical displacement field, $D$, independently of the band filling, $\nu$. At the same time, the flexibility of the TDBG platform leaves the apparent absence of superconductivity anywhere in its gate-tuned phase diagram particularly surprising and has led to speculation about a possible role of \ctwo symmetry in stabilizing the superconducting state in \m systems\cite{khalaf2021charged, park2022robust}. After initial reports of zero-resistance states in the TDBG conduction band\cite{liu2020tunable, shen2020correlated}, it is now widely believed that the low resistance observed in those experiments reflected reduced scattering due to broken spin or valley symmetries\cite{he2021symmetry}.

\begin{figure*}
  \includegraphics[width=1.0\textwidth]{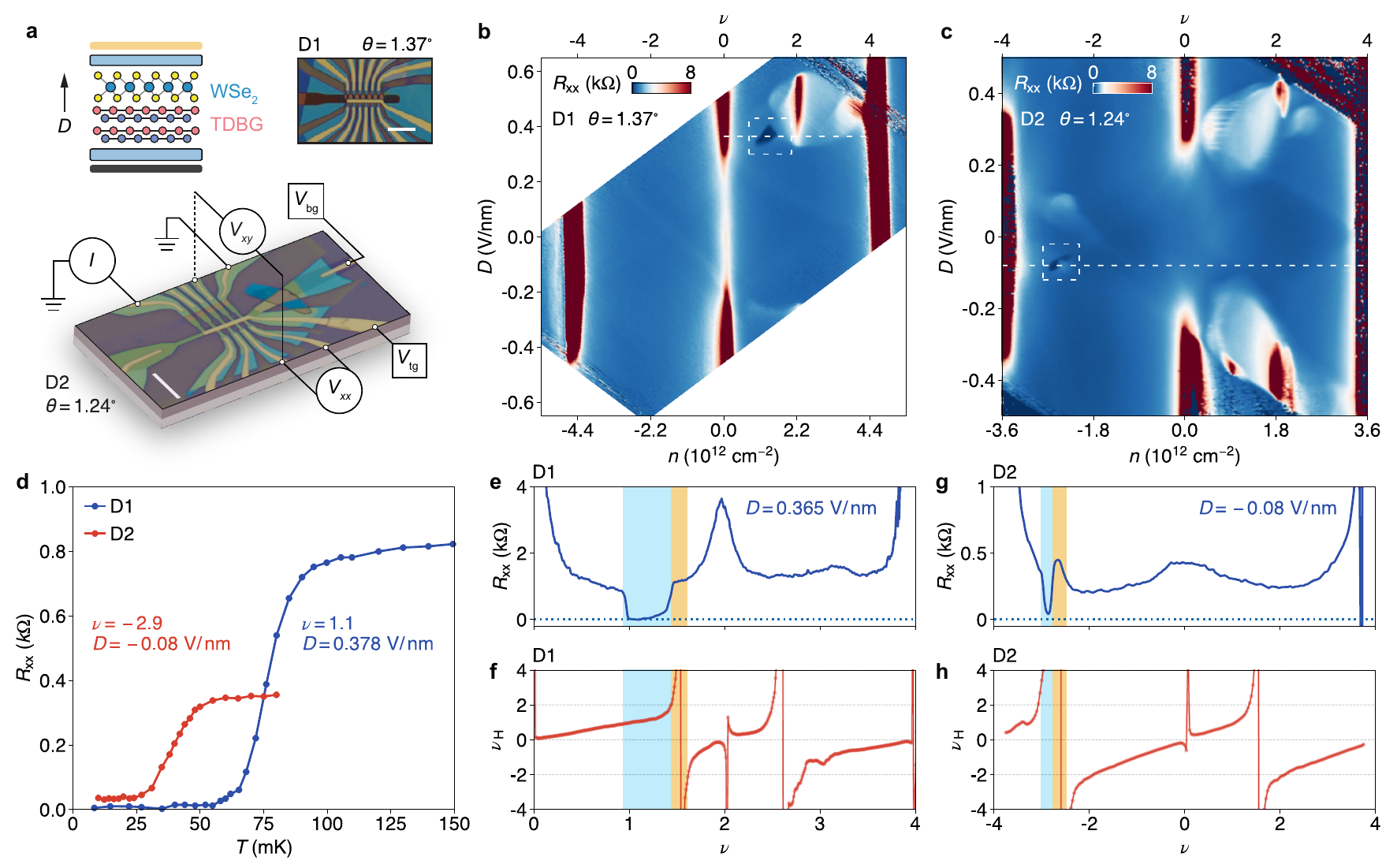}
  \caption{\label{fig:fig1}
  \textbf{Device characterization.} \textbf{a,} Schematic of the van der Waals stack, and optical images of D1 and D2, labeled with the measurement configuration. Top and bottom gates (gold and black) with hBN dielectric (blue) tune the displacement field $D$ and the carrier density $n$. \textbf{b, c,} Four-terminal resistance, $R_{xx}$, as a function of $n$ and $D$ for  D1(\textbf{b}) and D2(\textbf{c}) at zero magnetic field. The top axis indicates the moir\'e band filling factor $\nu$. Dashed boxes highlight localized pockets of superconductivity.
 \textbf{d,} Temperature dependence of $R_{xx}$ at specific values of $\{\nu,D\}$ in the superconducting pockets of D1 and D2. \textbf{e, f,} $\rxx(\nu)$ at $B=0$, and antisymmetrized Hall filling $\nu_H(\nu)$ at $B_{\perp} = \pm 0.8$ T, along the white dashed line in panel \textbf{b}. Superconductivity (shaded in blue) occurs immediately next to van Hove singularities, where $\nu_{H}$ diverges and changes sign (shaded in yellow). \textbf{g, h} Similar data for D2, obtained along the white dashed line in panel \textbf{c}.}
 \end{figure*}

 \begin{figure*}[ht]
    \includegraphics[width=1.0\textwidth]{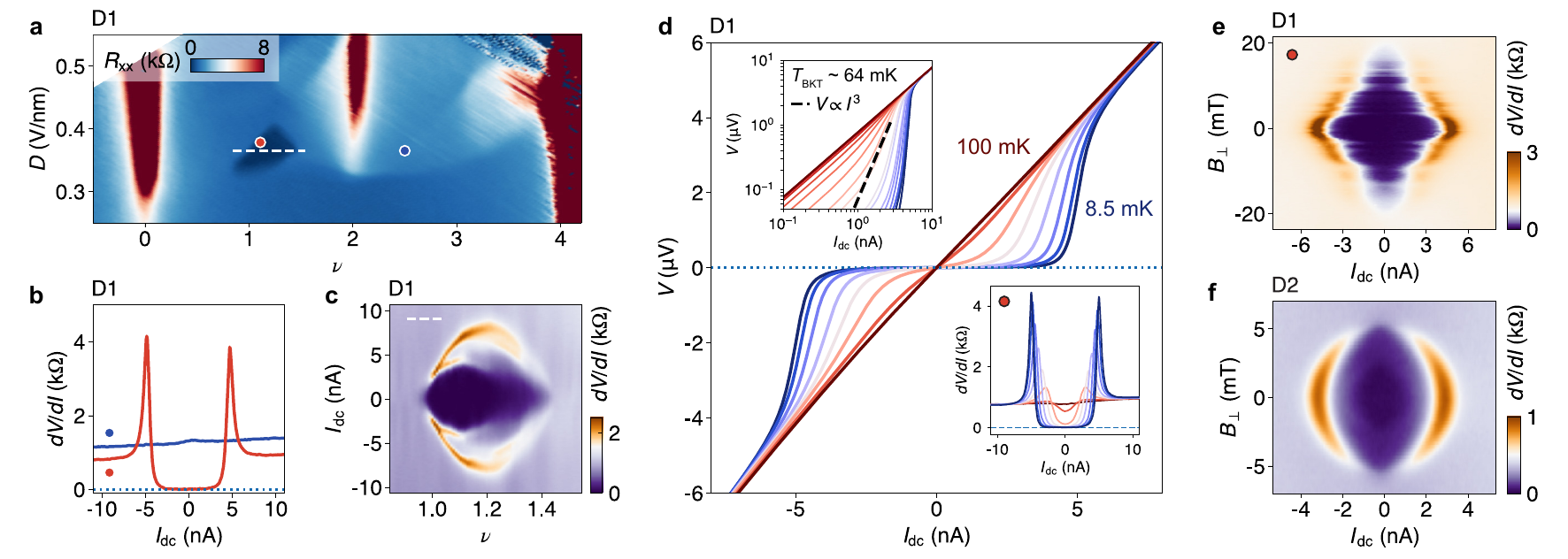}
    \caption{\label{fig:fig2}\textbf{Signatures of superconductivity.} \textbf{a,} $R_{xx}(\nu,D)$ showing the location of the low-resistance region in D1. \textbf{b,} Dependence of the differential resistance $dV/dI$ on $I_{dc}$ at the locations indicated with red and blue markers in panel \textbf{a}, showing sharp critical-current peaks inside the superconducting pocket but not outside. \textbf{c,} $dV/dI(I_{dc})$ along the dashed line in panel \textbf{a} ($D = 0.365$ V/nm), illustrating the evolution of the critical current with $\nu$. \textbf{d,} $I-V$ characteristics of superconductivity as a function of temperature, obtained by integrating $dV/dI(\idc)$ at the red marker in panel \textbf{a} (lower inset). $V(I_{dc})$ was used to determine the value $T_{\text{BKT}}$ of 64 mK (upper inset). \textbf{e, f,} Evolution of $dV/dI(\idc)$ with $\bperp$, showing Fraunhofer interference patterns in D1 (\textbf{e}, obtained at the red marker in panel \textbf{a}) but not in D2 (\textbf{f}, $\nu = -2.86$, $D = -0.086$ V/nm)}.
 \end{figure*}

Here, we show that TDBG bounded on one side by few-layer WSe$_2$ (Fig.~\ref{fig:fig1}a) does exhibit superconductivity, but the parameter range over which it is observed is qualitatively different than in TBG. We report data for samples with twist angles 1.37$^\circ$ and 1.24$^\circ$, referred to hereafter as D1 and D2. In D1, superconductivity is observed in a small gate-defined pocket close to filling $\nu\sim 1$ in the conduction band, near the so-called halo regions where broken-symmetry states are consistently reported\cite{he2021symmetry,kuiri2022}. In D2, the superconducting pocket is in the valence band near $\nu\sim-3$, where no correlation-induced modifications have been reported. In contrast to several other van der Waals structures, superconductivity in TDBG emerges in fully degenerate bands for both twist angles, but the superconducting pockets are localized to gate voltages that place the Fermi energy next to a van Hove singularity and close to phases with a tendency toward isospin (spin and/or valley) symmetry breaking.

Figure \ref{fig:fig1}a illustrates the sample structure, including TDBG and WSe$_2$ layers with bottom and top electrostatic gates. The devices were patterned into Hall bars, and lock-in measurements of longitudinal ($R_{xx}\equiv dV_{xx}/dI$) and transverse ($R_{xy}\equiv dV_{xy}/dI$) differential resistance were made in a dilution refrigerator, using top gate ($V_{tg}$) and back gate ($V_{bg}$) voltages to set $\nu$ and $D$. Except where noted otherwise, measurements are made at a base temperature of 10 mK. The resulting resistivity maps are shown in Fig.~\ref{fig:fig1}b and c for D1 and D2 respectively, and are broadly consistent with published data for TDBG samples with twist angles ranging from $1.2$ to $1.3^\circ$\cite{shen2020correlated,liu2020tunable, caoTunableCorrelatedStates2020,he2021symmetry,kuiri2022}. In particular, halo regions of higher resistivity (lighter blue) appear in the conduction band, bisected by additional insulating states at $\nu=2$, and faint diagonal features cross at $D=0$ around $\nu=-2$ in the valence band\cite{he2021symmetry}.

Also visible in Fig.~\ref{fig:fig1}b and c are small dark blue pockets, where the resistance drops close to zero from a smooth background of hundreds of ohms. The low resistance state appears below 80 mK in D1, where $R_{xx}$ drops to within experimental uncertainty of zero, or 40 mK in D2 where $R_{xx}$ drops to a few tens of ohms (Fig.~\ref{fig:fig1}d). The superconducting pockets in both samples are observed only for the direction of applied $D$ that is expected to pull the relevant \m conduction (D1) or valence (D2) band closer to the WSe$_2$\cite{liu2022ferromagnet, zhang2022spin}, where the direction of positive $D$ is defined in Fig.~\ref{fig:fig1}a. 

Comparing $B=0$ linecuts of the $R_{xx}$ data to Hall measurements obtained with out-of-plane magnetic field $\bperp=0.8$ T (Figs.~\ref{fig:fig1}e--h), the gate-voltage locations of the low-resistance features are adjacent to van Hove singularities where the Hall coefficient, $R_H\equiv R_{xy}/|\bperp|$, goes through zero and the Hall density, $n_H\equiv 1/(eR_H)$, diverges. For easier interpretation, we report here the Hall filling, $\nu_H\equiv 4n_H/n_s$, reflecting the Hall density normalized by the measured carrier density of a fully filled \m band, $n_s$, with a factor of four to account for spin and valley degeneracy (Figs.~\ref{fig:fig1}f,h).

Figure \ref{fig:fig2} illustrates the fragility of the low-resistance state to direct current, \idc, and to $\bperp$, confirming the presence of superconductivity. We focus primarily on the stronger state in D1 (Fig.~\ref{fig:fig2}a). Figure~\ref{fig:fig2}b and c contrast the $I_{dc}$ breakdown of the $R_{xx}$=0 state within the superconducting pocket, displaying sharp peaks in $dV/dI$ at a nanoampere-scale critical current, with the nearly flat $dV/dI(I_{dc})$ trace at nearby locations in $\{\nu,D\}$.  The temperature dependence of $V(I_{dc})$ traces, obtained by integrating $dV/dI(I_{dc})$, shows the classic evolution of 2D superconductors from step-like transitions at low temperature to ohmic dependence above 100 mK (Fig.~\ref{fig:fig2}d). At this specific $\{\nu,D\}$ setting, a Berezinskii--Kosterlitz--Thouless (BKT) analysis of the evolution of the $V \propto I_{dc}^{\alpha}$ power law near the critical current indicates a BKT transition temperature of $64\pm 5$ mK where $V \propto I_{dc}^{3}$ (Fig.~\ref{fig:fig2}d upper inset).

The clearest demonstration of superconductivity is found in the $\bperp$ dependence of $dV/dI(\idc)$ for D1 (Fig.~\ref{fig:fig2}e). The repeated collapse and revival of the critical current with \bperp, also referred to as a Fraunhofer pattern, results from SQUID-like quantum interference of transport through Josephson junctions and is commonly used as a confirmation of superconductivity in TBG experiments \cite{arora2020superconductivity,cao2018unconventional}. The characteristic field scale of the fluctuations, $\Delta \bperp=$ 2--3 mT, is consistent with a domain size $\sqrt{\Phi_0/\Delta \bperp}=0.7~\mu$m that is close to the Hall-bar width of 1 $\mu$m. Equivalent data for D2 (Fig.~\ref{fig:fig2}f) lacks the Fraunhofer modulations of D1, tentatively explained by the weaker superconductivity and the faster collapse of the critical current with \bperp, or by a lack of Josephson junctions in D2.

 \begin{figure*}
    \includegraphics[width=1.0\textwidth]{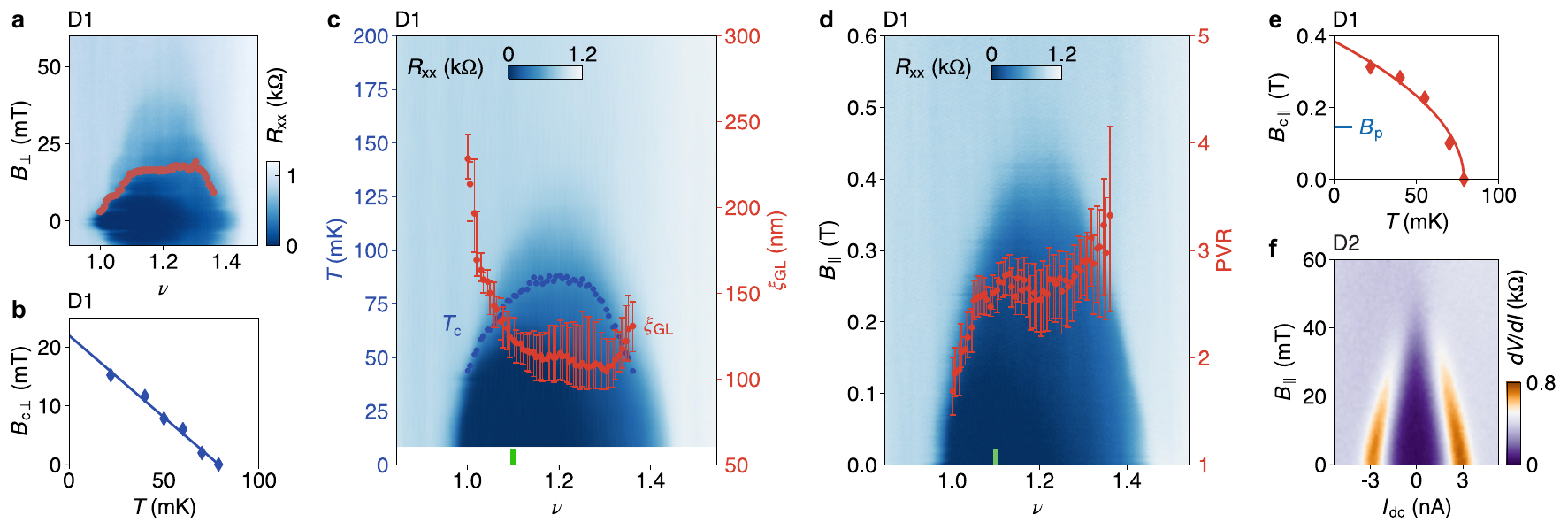}
    \caption{\label{fig:fig3}\textbf{Metrics of superconductivity in TDBG.} Panels \textbf{a--e} were obtained at a fixed $D = 0.365$ V/nm, along the white dashed line in Fig.~\ref{fig:fig2}a. \textbf{a,} $\bperp$ dependence of $R_{xx}$, showing Fraunhofer modulations. The red markers indicate the critical field, $B_{c \perp}$, defined by a threshold of half of its normal-state resistance. \textbf{b,} Temperature dependence of $B_{c \perp}$ obtained at $\nu = 1.1$. The straight-line fit was used to extract $B_{c\perp}^0 = 22$ mT. \textbf{c,} Temperature dependence of $\rxx$, used to determine the $T_c$, shown together with Ginzburg--Landau coherence length, $\xi_{\text{GL}}$, calculated using $B_{c\perp}^0$. Error bars show $\xi_{\text{GL}}$ extracted using 40\% and 60\% thresholds for $B_{c\perp}$. 
    \textbf{d,} \bpar~dependence of $\rxx$, together with the Pauli-limit-violation ratio (PVR) calculated using $B_{c\parallel}^0$. Error bars show PVR calculated using 60\% and 40\% thresholds for $B_{c\parallel}$. \textbf{e,} Temperature dependence of $B_{c\parallel}$ obtained at $\nu = 1.1$. The blue marker shows the  Pauli-limit field. \textbf{f,} Suppression of non-linear $dV/dI$ by $\bpar$ for D2 ($\nu = -2.86$, $D = -0.08\text{ V/nm})$, where $B_P=77$ mT}
 \end{figure*}

In Fig.~\ref{fig:fig3} we show how the metrics of superconductivity vary with the band filling across the pocket in D1, following the line cut marked in Fig.~\ref{fig:fig2}a. At each value of $\nu$, the collapse of superconductivity with $\bperp$ (Fig.~\ref{fig:fig3}a) can be used to determine a critical out-of-plane magnetic field, $B_{c\perp}$, defined as the field where the resistance rises to one half of its normal-state value. Measuring $B_{c\perp}$ as a function of temperature enables an extrapolation to zero temperature (Fig.~\ref{fig:fig3}b), giving $B_{c\perp}^0$ from which the Ginzburg--Landau coherence length, $\xi_{\text{GL}}=\sqrt{\Phi_{0}/(2\pi B_{c \perp}^0)}$, is determined\cite{supplement}. $\xi_{\text{GL}}$ is around 100 nm across most of the superconducting dome, comparable to the mean free path, $\ell_m\sim 200$ nm, which can be estimated from the onset of Shubnikov--de Haas oscillations around 400 mT\cite{supplement}. For D2, a similar analysis yields $\xi_{GL}\sim$250 nm and $\ell_m\sim 600$ nm.

Consistent with the two-dimensional nature of the superconductivity in TDBG, the in-plane critical fields $B_{c\parallel}$ are much greater than the out-of-plane $B_{c\perp}$ for both samples. Conventional superconductivity, with spin-singlet Cooper pairs, is destroyed when the Zeeman energy induced by the in-plane magnetic field exceeds the superconducting gap, leading to the Pauli (Chandrasekhar--Clogston) limit field, $B_{p}=1.76 k_B T_c^{0} g^{-1/2}/\mu_B$, above which superconductivity is expected to vanish ($g$ is the Land\'e $g$-factor and $T_c^0$ is the critical temperature at $B=0$)\cite{PhysRevLett.9.266,tinkham2004introduction}.

The superconducting states in D1 and D2 are very different in their resilience to $\bpar$, with D1 exceeding the Pauli limit by a factor of 2--3 but D2 remaining under the Pauli limit. Figure~\ref{fig:fig3}d shows the collapse of superconductivity with $\bpar$ for D1, over the same range of $\nu$ mapped out in Figs.~\ref{fig:fig3}a,c. For quantitative analysis, $B_{c\parallel}$ is defined by where $R_{xx}$ reaches a threshold of half of its normal-state resistance. The temperature dependence of $B_{c\parallel}$ can be fitted to the phenomenological relation $T/T_{c}^{0} = 1 - (B_{c\parallel}/B_{c\parallel}^{0})^{2}$ commonly used for Pauli-limited superconductivity\cite{cao2018unconventional, zhou2021superconductivity, zhang2022spin, cao2021pauli} (see for example Fig.~\ref{fig:fig3}e), and the resulting $B_{c\parallel}^0$ compared to the value of $B_P$ assuming $g=2$.
The ratio, $\text{PVR}\equiv B_{c\parallel}^0/B_P$, is between 2 and 3 across most of the dome (Fig.~\ref{fig:fig3}d) for D1, but is less than 1 for D2 at optimal doping (Fig.~\ref{fig:fig3}f)\cite{supplement}. Recent studies of superconductivity in the family of magic-angle twisted graphene monolayers have identified a similar resilience against pair-breaking by in-plane magnetic field, which depended  strongly  on the number of layers \cite{cao2018unconventional,park2022robust, cao2021pauli, zhang2022promotion}.

\begin{figure*}
    \includegraphics[width=1.0\textwidth]{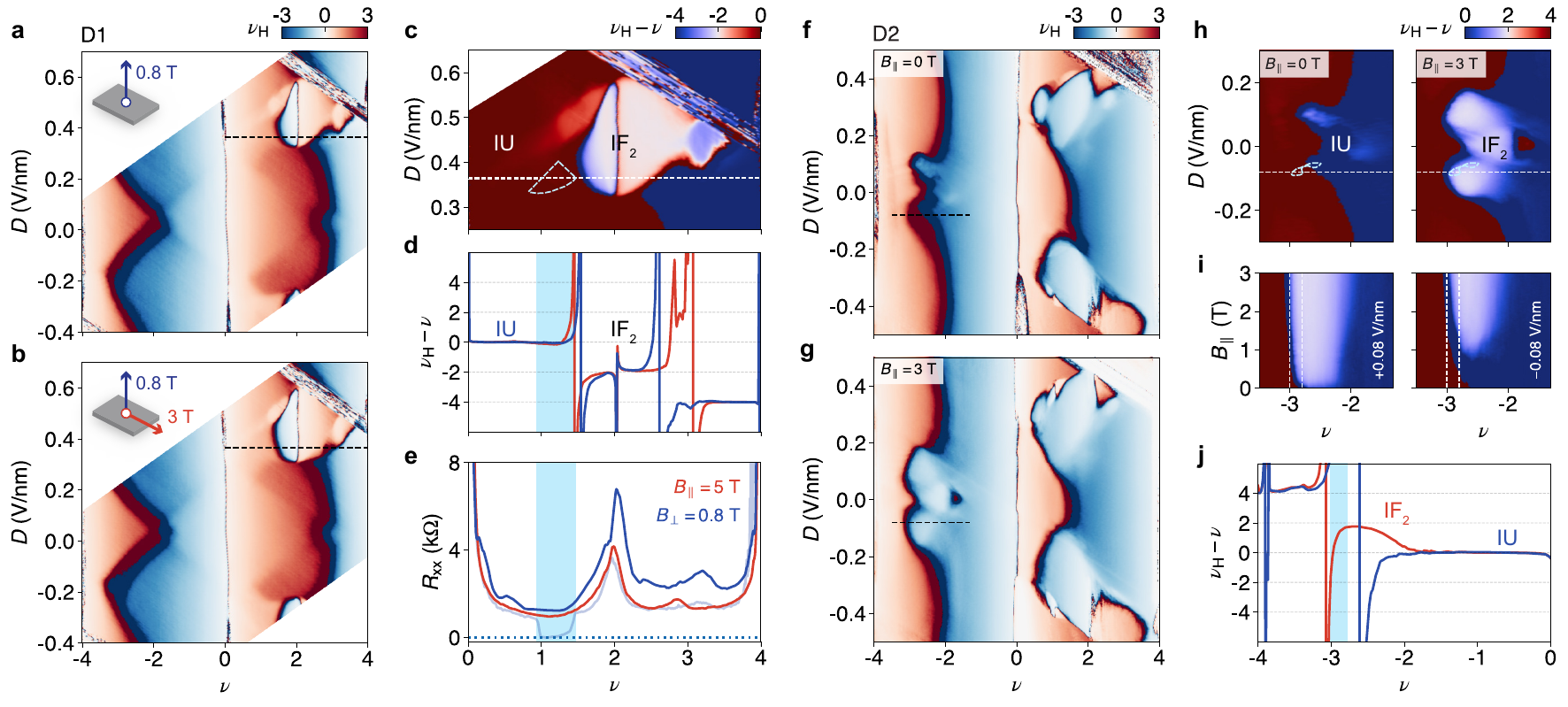}
    \caption{\label{fig:fig4}
    \textbf{Van Hove singularities and isospin polarized phases.} \textbf{a,} Antisymmetrized $\nu_{H}(\nu, D)$ for D1 ($\bperp= \pm 0.8$ T). \textbf{b,} Same as \textbf{a}, but with an additional $\bpar=3$ T. \textbf{c, } Subtracted Hall filling, $\nu_{H} - \nu$, near the upper halo region in panel \textbf{a}, highlighting the IF$_2$ state. \textbf{d, }Line cuts of $\nu_{H} - \nu$ along $D = 0.365$ V/nm at $\bpar=0$ (blue) and $\bpar=3$ T (red), showing unpolarized (IU) regions with degeneracy 4 as well as broken-symmetry regions with degeneracy 2 (IF$_2$). \textbf{e, }Effects of $B_{\perp}$ and $B_{\parallel}$ independently on the $\nu = 2$ insulating state, along $D = 0.365$ V/nm. \textbf{f, g,} Analogous measurements to \textbf{a, b} for D2, showing the $\bpar$-dependent features in the valence band where superconductivity emerges. \textbf{h,} $\nu_{H} - \nu$ for the valence-band features from panels \textbf{f} and \textbf{g}. \textbf{i, }Emergence of the IF$_2$ regions with $\bpar$, for $D=+0.08$ and $D=-0.08$ V/nm. \textbf{j,} Line cuts through the data in panel \textbf{h}.}
 \end{figure*}

The conditions under which superconductivity appears in TDBG offer an insightful comparison with other graphene systems, with or without a twist. An apparently unifying feature of graphene-based superconductivity in all known systems is the proximity to symmetry-broken phases. In most systems---\m stacks, BBG, and the SC2 phase of RTG---superconductivity appears to emerge directly out of a phase in which two of the four degeneracies were lifted, as determined by magnetoresistance measurements in the normal phase, once superconductivity has been suppressed. Figure~4 summarizes magnetoresistance measurements in TDBG, indicating that superconductivity in this system emerges out of fully degenerate bands, adjacent to regions with a tendency toward isospin ferromagnetism.  

Starting from four spin- and valley-degenerate bands, the Hall filling $\nu_H$ departs from the total filling $\nu$ by $\pm$1, $\pm$2, or $\pm$3 when isospin-symmetry breaking leaves some bands completely filled (therefore not contributing to $\nu_H$), so the subtracted Hall filling $\nu_H-\nu$ probes band polarization. Figure~\ref{fig:fig4}a and b show $\nu_H$ for D1 at $\bpar=0$ and $\bpar=3$ T, respectively, with $\delnu$ in the vicinity of the superconducting pocket highlighted in Fig.~\ref{fig:fig4}c.  Line cuts through the data in Fig.~\ref{fig:fig4}c, and analogous data at 3 T (see Fig.~\ref{fig:fig4}b) are shown in Fig.~\ref{fig:fig4}d. The pocket where superconductivity appears, indicated in blue in Fig.~\ref{fig:fig4}d, is  isospin unpolarized (IU, $\delnu=0$), but the halo  immediately adjacent to the superconducting region is an isospin ferromagnet with degeneracy 2 (IF$_2$, $\delnu=-2$). 

In order to infer the broken symmetry of the IF$_2$ halo, we studied the effect of perpendicular and in-plane magnetic fields on the resistance of the correlated insulator at half filling (Fig.~\ref{fig:fig4}e). In contrast to previous reports of spin polarization in the halo regions of TDBG\cite{liu2020tunable, caoTunableCorrelatedStates2020, shen2020correlated}, the IF$_2$ halo in D1 appears not to be spin-polarized, as $\rxx$ at the insulating $\nu=2$ peak in the middle of the halo does not rise with in-plane magnetic field up to 5 T. The halo regions in D2 do appear to be spin polarized, extending significantly in gate voltage when $\bpar$ is raised to 3 T and with a $\nu=2$ peak that rises dramatically with in-plane field\cite{supplement}.  The fact that superconductivity does not appear next to the halo region in D2 might be due to the different isospin symmetries that are broken there, or to the fact that symmetry breaking is stronger in the D2 halo, with the IF$_2$ region extending down to $\nu=0.5$ in D2 but only to $\nu=1.5$ in D1.

Another contrast between D1 and D2 is the change of $\nu_H$ with $\bpar$ in the valence band of D2 (Fig.~\ref{fig:fig4}f,g), whereas $\nu_H$ is unaffected by $\bpar$ in the valence band of D1 (Fig.~\ref{fig:fig4}f,g). Examining $\nu_H-\nu$ in the D2 valence band more closely (Fig.~\ref{fig:fig4}h), it is apparent that three lobes of IF$_2$ isospin ferromagnet  appear at $\bpar$=3 T, with hints of this structure visible at $\bpar$=0. The emergence of the IF$_2$ state with $\bpar$ is mapped out in Fig.~\ref{fig:fig4}i, for line cuts at $D=+0.08$ V/nm and $-0.08$ V/nm; the state appears rapidly with $\bpar$ at $D=+0.08$ V/nm, but slowly at $D=-0.08$ V/nm, where superconductivity is observed.
 
The need for higher $\bpar$ to form the isospin ferromagnetic phase at $D=-0.08$ V/nm, compared to the phase at $D=+0.08$ V/nm, seems to imply a weakening of the symmetry-broken state by the WSe$_2$. This observation is reminiscent of the stabilization of superconductivity across a wider region of the phase diagram in BBG due to adjacent WSe$_2$ \cite{zhang2022spin}, where the BBG would otherwise break into multiple isospin-polarized phases \cite{zhou2022isospin}.

Taken together, our measurements demonstrate superconductivity enhanced in proximity to broken-degeneracy phases, but suppressed when degeneracy is fully broken. For D1, the nearby IF$_2$ state is reached by a small change in gate voltage, whereas in D2 the nearby IF$_2$ state is reached by adding Zeeman energy via $\bpar$. The fact that both the IF$_2$ states and the superconducting pockets are localized to small fractions of the phase diagram, yet the two are consistently next to each other, offers strong evidence of a correlation between them. It does not, however, rule out a scenario in which the same characteristics of the gate-tuned band structure (such as the diverging density of states at the van Hove singularity) lead to both effects, with the tendency to isospin ferromagnetism overwhelming singlet pairing for superconductivity when it is sufficiently strong.

The role of WSe$_2$ is clear from the emergence of superconductivity only for the sign of D that pulls the relevant band into stronger proximity with the transition metal dichalcogenide. What precisely that role is, remains less clear. As a test for proximity-induced spin--orbit coupling, we searched for weak (anti-)localization corrections to the resistivity in both samples. Only weak localization was observed in D2, and only in isolated regions of the phase diagram\cite{supplement}; neither weak localization nor anti-localization was observed in D1. These observations call for further investigation, but cannot conclusively identify spin--orbit interaction as playing an important role in TDBG superconductivity. An intriguing possibility is the proposal in ref.~\cite{chou2022enhanced} that virtual tunneling into WSe$_2$ reduces the native Coulomb repulsion between electrons in a Cooper pair, enabling other pairing mechanisms to stabilize superconductivity.

Finally, we turn to possible interpretations of the Pauli violation for D1, but none for D2. One possibility is that proximity-induced spin--orbit interaction in D1 raises the Pauli limit for that sample\cite{tinkham2004introduction, klemm1975theory} but not for D2. The lack of weak anti-localization in either sample makes this explanation less likely, although a pure Ising spin--orbit interaction would not lead to anti-localization\cite{mccann2012z,wakamura2019spin}. Another possibility is that the valence band superconductivity in D2 has conventional spin-singlet Cooper pairs (therefore adhering to the Pauli limit with $g=2$), whereas superconductivity in D1 has valley-singlet pairing, and the effective $g$-factor splitting the valley pairs due to an in-plane field might be significantly less than 2\cite{qin2021plane}. This explanation is consistent with the spin (valley) isospin polarizations for nearby regions in D2 (D1), providing further support for the possibility of isospin fluctuations as a mechanism for pairing in both samples\cite{liu2022isospin, qin2021plane, huang2022pseudospin}.

\subsection{Acknowledgements} 
We thank Allan MacDonald, Jihang Zhu, Nemin Wei, Sankar Das Sarma, Yang-Zhi Chou, Andrew Potter, and Marcel Franz for fruitful discussions. MK acknowledges a postdoctoral research fellowship from Stewart Blusson Quantum
Matter Institute (SBQMI). Experiments at UBC were undertaken with support from SBQMI, the Natural Sciences and Engineering Research Council of Canada, the Canada Foundation for Innovation, the Canadian Institute for Advanced Research, and the Canada First Research Excellence Fund, Quantum Materials and Future Technologies Program, and ERC Synergy
funding for Project 941541. K.W. and T.T. acknowledge support from JSPS KAKENHI (Grant
Numbers 19H05790, 20H00354 and 21H05233).

\subsection{Author Contribution}
R.S. fabricated the devices, with help from M.K.; R.S.
performed measurements; R.S., M.K., and
J.F. interpreted the data and wrote the manuscript; J.F. supervised the experiment.
K.W. and T.T. provided the hBN crystals.

\onecolumngrid
\newpage
\thispagestyle{empty}
\mbox{}


\section*{Supplementary material}

# Supplementary Material for "Superconductivity in Twisted Double Bilayer Graphene Stabilized by WSe$_2$"

Ruiheng Su[1], Manabendra Kuiri[1], Kenji Watanabe[2], Takashi Taniguchi[3], and Joshua Folk[1]*

[1] *Department of Physics and Astronomy & Stewart Blusson Quantum Matter Institute,*
*University of British Columbia, Vancouver BC, Canada V6T 1Z4*
[2] *Research Center for Functional Materials, National Institute for Materials Science,*
*Namiki 1-1, Tsukuba, Ibaraki 305-0044, Japan and*
[3] *International Center for Materials Nanoarchitectonics,*
*National Institute for Materials Science, Namiki 1-1, Tsukuba, Ibaraki 305-0044, Japan*

**This PDF file includes:**
Materials and Methods
Supplementary Figs. S1 to S21
References

---

* jfolk@physics.ubc.ca

## I. MATERIALS AND METHODS

We assembled our WSe$_2$, AB-AB stacked twisted bilayer graphene heterostructure using the cut-and-stack method. We first mechanically exfoliated hexagonal boron nitride, few-layer tungsten diselenide, and graphene flakes onto clean SiO$_2$ (300 nm) /Si chips. Using optical contrast, we identified large-area bilayer graphene flakes. The bilayer graphene flake was cut into halves using a sharp ∼1 $\mu$m diameter tungsten dissecting needle prior to dry transfer [1]. We used a stamp made of a poly(bisphenol A carbonate) (PC) film laminated over a polydimethylsiloxane hemisphere [2] on a glass slide attached to a micromanipulator. The top hBN, WSe$_2$ were picked up at 90 °C. Next, the first half of the bilayer graphene flake was picked at 50 °C, and the stage was rotated by $\theta = 1.4°$, slightly overshooting to achieve the target angle. The remaining half of the bilayer graphene flake was picked up at the same temperature. The remainder of the stack - bottom hBN and graphite bottom gate - were picked up at 100°C. The stack was dropped onto a clean SiO$_2$/Si substrate with pre-patterned Ti/Au alignment marks for further processing. The top gate was first defined on the stack using electron beam lithography (EBL), followed by electron beam evaporation of Cr/Au (5 nm/50 nm). Edge contacts were made by dry etching the stack using a CHF$_3$/O$_2$ plasma (40 sccm/4 sccm, 4 pa chamber pressure, 60 W RF power), followed by evaporation of Cr/Au (8 nm/70 nm). The device was shaped into a Hall bar using a final CHF$_3$/O$_2$ plasma etch. During the metalization, we ensured a chamber pressure of at least $5 \times 10^{-7}$ torr, and chromium and gold were evaporated at a rate of 0.5 and 1.0 Å/s, respectively.

### A. Transport Measurements

We performed the transport measurements in a Bluefors LD dilution refrigerator with a base temperature of 10 mK, equipped with a 3-axis vector magnet, and using lock-in amplifiers synchronized to 17 Hz to reduce the measurement noise. Measurements in the superconducting state used an AC bias current of 0.1 nA, while measurements in a finite magnetic field used an AC current bias of 5 nA unless otherwise stated. All measurements were performed at the base temperature, and the tilt angle between the normal of the sample plane and the out-of-plane axis of the vector magnetic was kept $< 1°$. Top and bottom gate voltages were used to independently tune the carrier density, $n$, and electric displacement field, $D$, according to the equations: $n = (C_{bg}V_{bg} + C_{tg}V_{tg})/e$ and $D = (C_{bg}V_{bg} - C_{tg}V_{tg})/2\epsilon_0$, where $C_{bg}$ and $C_{tg}$ are respectively the bottom and top gate capacitance per unit area, $e$ is the elementary charge and $\epsilon_0$ is the vacuum permittivity. $C_{bg}, C_{tg}$ using the slope of the hall density, $n_H = 1/eR_{xy}$, where $R_{xy} = (R_{xy}[B_\perp > 0] - R_{xy}[B_\perp < 0])/(2|B|)$ is the antisymmetrized Hall resistance, and $e$ is the electron charge. The gate capacitance found for devices D1 and D1 is shown in Fig. S6.

We estimated the device twist angle, $\theta$, from the density corresponding to full filling of the miniband $n(\nu = \pm 4)$ = $8\theta^2/\sqrt{3}a^2$, where $a = 0.246$ nm is the lattice constant of graphene. In our device, $\nu = \pm 4$ corresponds $n = \pm 3.6 \times 10^{12}$cm$^{-2}$ and $n = \pm 4.4 \times 10^{12}$ cm$^{-2}$, which translates to twist angles $\theta \approx 1.24°$ and $\theta \approx 1.37°$ for devices D1 and D2 respectively.

### B. Critical current measurements

We used a lock-in technique to directly measure the differential resistance, $dV/dI$, as a function of the applied DC current. This was done by measuring the AC response of the device to a signal consisting of a DC current with a small AC component. The DC current was sourced from a home-built digital-to-analog converter with a 1 MΩ bias resistor. The AC current was sourced from a lock-in amplifier through the same 1 MΩ bias resistor. The circuit used is shown in Fig. S7. The current through the device was amplified by a low-noise current amplifier and the AC component, $\Delta I = I_{ac}$, was directly measured by a lock-in amplifier. The AC voltage drop between two adjacent contacts, $\Delta V = V_{ac}$, was amplified by a low-noise differential voltage amplifier and measured by another lock-in amplifier. The differential resistance is then given by the ratio $dV/dI = \Delta V/\Delta I$. We used an AC current bias of 0.1 nA to avoid inadvertent broadening of sharp features in $dV/dI$ as a function of $I_{dc}$.

### C. Critical temperature and magnetic fields

We fit $r(z) = Az + C$ to the portion of $R_{xx}(z)$ (where $z$ represents the temperature, $T$, or applied magnetic field $B$) when the superconducting sample returns its normal state. We then found the $z$ value for which $\alpha \times r(z)$ intersects $R_{xx}(z)$, where $\alpha$ is the percentage resistance threshold, to determine the critical temperature and magnetic fields, $T_c^\alpha$, $B_c^\alpha$. A 50% threshold was used unless otherwise specified.

## II. SUPPLEMENTARY FIGURES

### A. Device Details

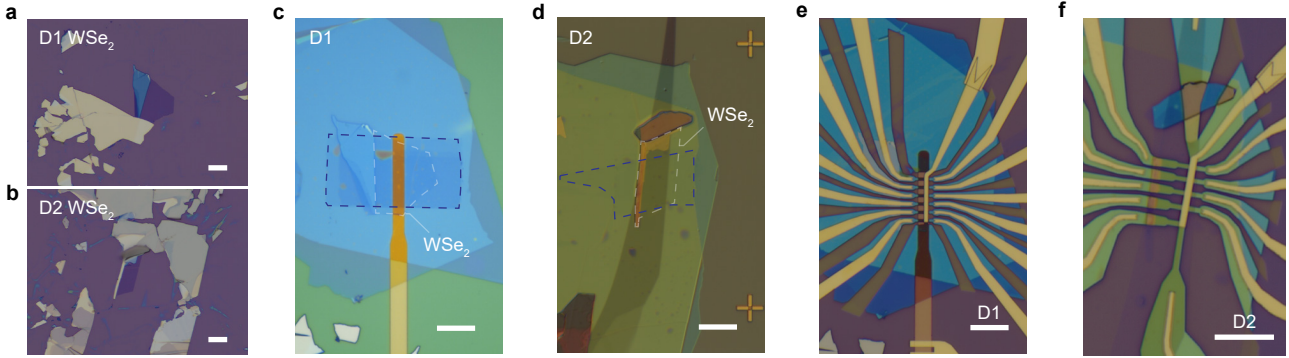

FIG. S1. **Device Fabrication. a, b,** Optical images of the mechanically exfoliated few-layer $WSe_2$ on $Si/SiO_2$ substrate with 285 nm of oxide, used in device D1 and D2, respectively. **c,** Optical image of the stack consisting of $hBN/WSe_2/TDBG/hBN/gold$ (from top to bottom), device D1. **d** Optical image of the stack, D2 consisting of $hBN/WSe_2/TDBG/hBN/graphite$. The $WSe_2$(TDBG) is outlined in white (blue) dashed lines. **e** Completed device, D1 with contacts, patterned into a Hall bar geometry. Here, top and bottom gates are made of gold (Au). **f** Completed Device, D2. Here graphite serves as the bottom gate, and gold (Au) serves as the top gate. In all panels, the scale bar is 10 $\mu$m.

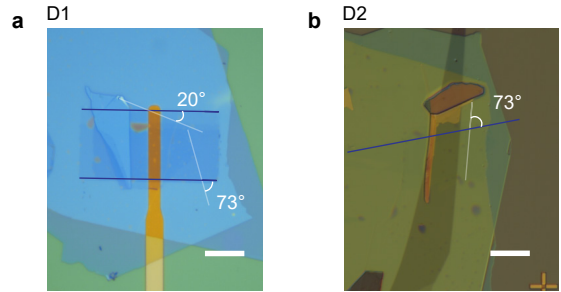

FIG. S2. **Misalignment between $WSe_2$ and TDBG.** Optical images for device D1 (**a**), and device D2 (**b**), before deposition of edge contacts. The misalignment angle between the TDBG and $WSe_2$ was estimated using the angles between potential straight crystal edges of the TDBG and $WSe_2$ flakes. For D1, we estimate the misalignment to be between 13° and 20°. For D2, we estimate a misalignment angle of 13°.

## B. Twist angle characterization at 4 K

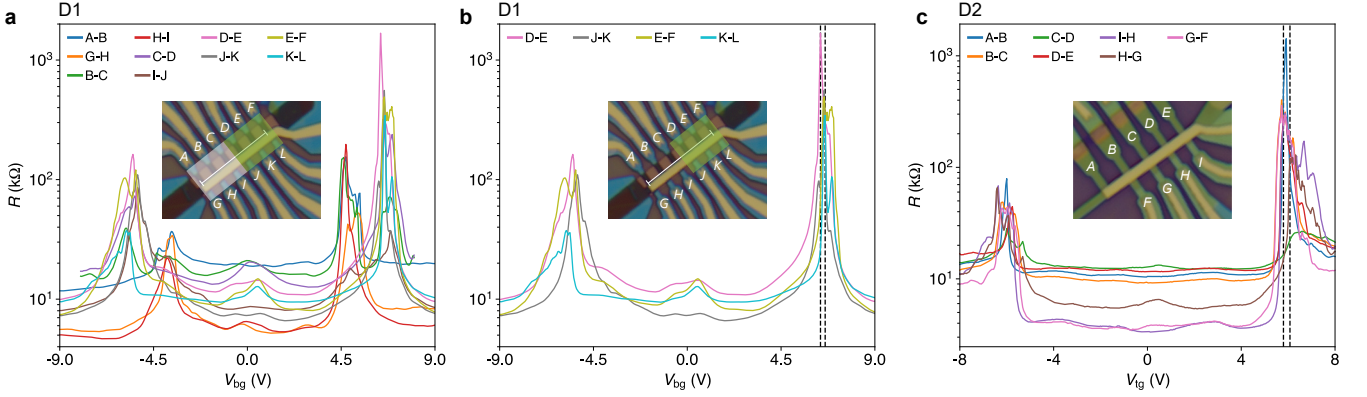

FIG. S3. **Characterization of twist angle homogeneity. a,** Two probe resistance as a function of bottom gate voltage between neighboring contact pairs at $T = 4$ K, for device D1. The inset shows the optical image of the device with voltage pairs labeled. The device shows two contrasting twist angles along the channel of the hall bar, shaded by white and green. **b,** Two probe resistance for the contact pairs in the device region shaded in green. We used offset between peaks (highlighted by black dashed lines), $\delta V$, at full band filling to estimate the local twist angle variation. The carrier density variation, $n(V_{tg} = 0, \delta V_{bg})$, is $\sim 14 \times 10^{10}$cm$^{-2}$, which translates an angle variation of $\sim 0.02°$ over $\sim 4.5$ $\mu$m. All subsequent measurements and the observation of superconductivity were made using these contact pairs. Four terminal resistance data presented in the main manuscript used contacts K-L as the voltage probes, while hall measurements used contacts E-K. **c,** Two terminal resistance for all contact pairs in device D2 at $T = 4$ K. Similar to **b**, we used $n(\delta V_{tg}, V_{bg} = 0) \sim 15^{10}$cm$^{-2}$ to estimate an angle variation of $\sim 0.03°$ over 10 $\mu$m. Longitudinal resistance data presented in the main manuscript used contact I-H as the voltage probes, while hall resistance data used contacts D-H as the voltage probes.

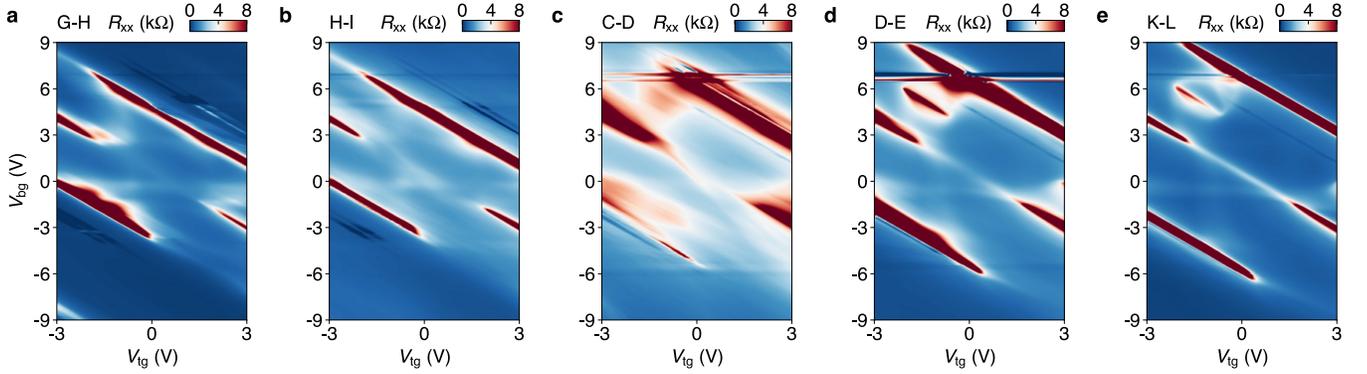

FIG. S4. **Four terminal measurements for D1 at $T = 4$ K. a-e,** Color plot of the four terminal resistance as a function of top and bottom gate voltages for the different contact pairs shown in Fig. S3a, showing the evolution of the $R_{xx}$ peaks at full band filling over the length of the device, $T = 4$ K. The labels on the top left corner of each panel indicate the contacts used as the voltage probes in the four-terminal measurements. Data were acquired simultaneously by multiple lock-in amplifiers. In these measurements, contacts A and F located at the end of the hall bar were used as the source and drain contacts. Dark blue features parallel to full-filling resistance peaks are due to twist angle variation, which was absent in subsequent measurements after changing the source and drain contacts to contacts C and F to isolate regions of the device with a uniform twist angle (green shaded region in S3b). Due to dielectric leakage, $V_{tg} > \pm 3$ V were inaccessible during the experiment.

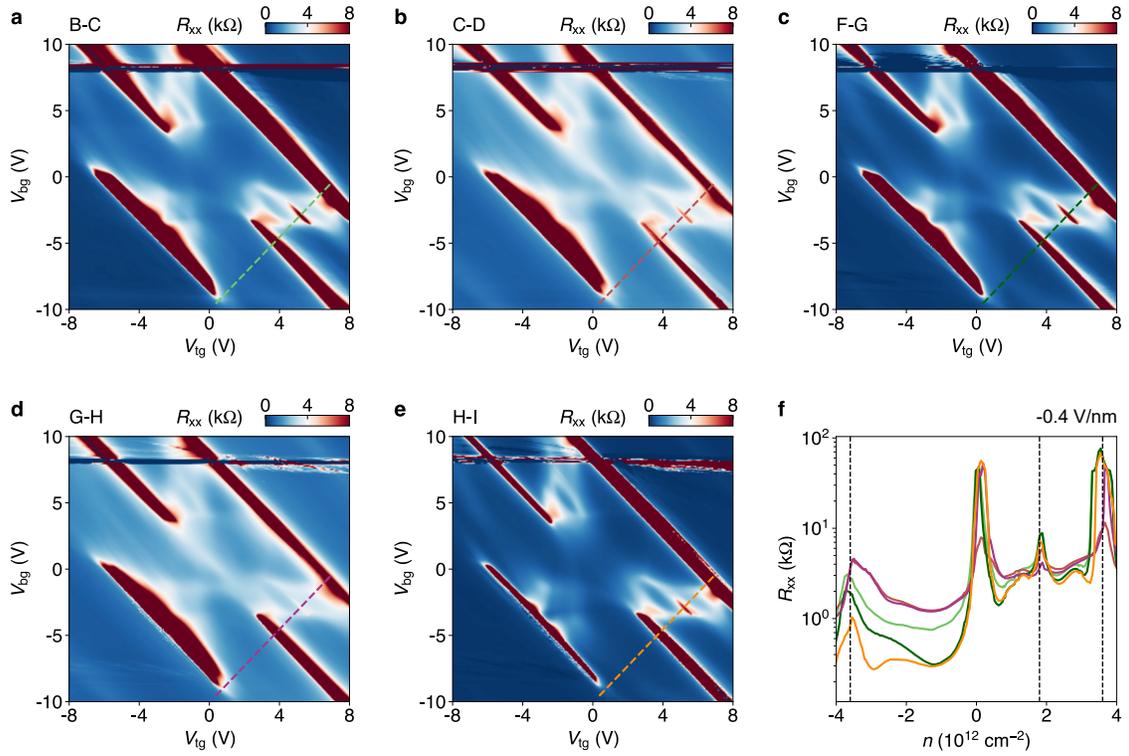

FIG. S5. **Four terminal measurements for D2 at** $T = 4$ **K. a-e** Shows the four terminal color plot of the measured resistance, $R_{xx}$ as a function of top and bottom gate voltages for the different contact pairs, shown in Fig. S3c, $T = 4$ K. **f,** Shows the line cuts of $R_{xx}$ along the color dashed lines in **a-e** were $D = -0.4$ V/nm. The high twist angle homogeneity in D2 is evident in the close alignment of resistance peaks at carrier densities $n = 1.8 \times 10^{12}$ cm$^{-2}$ ($\nu = 2$) and $\pm 3.6 \times 10^{12}$ cm$^{-2}$ ($\nu = \pm 4$) across all contact pairs along the device.

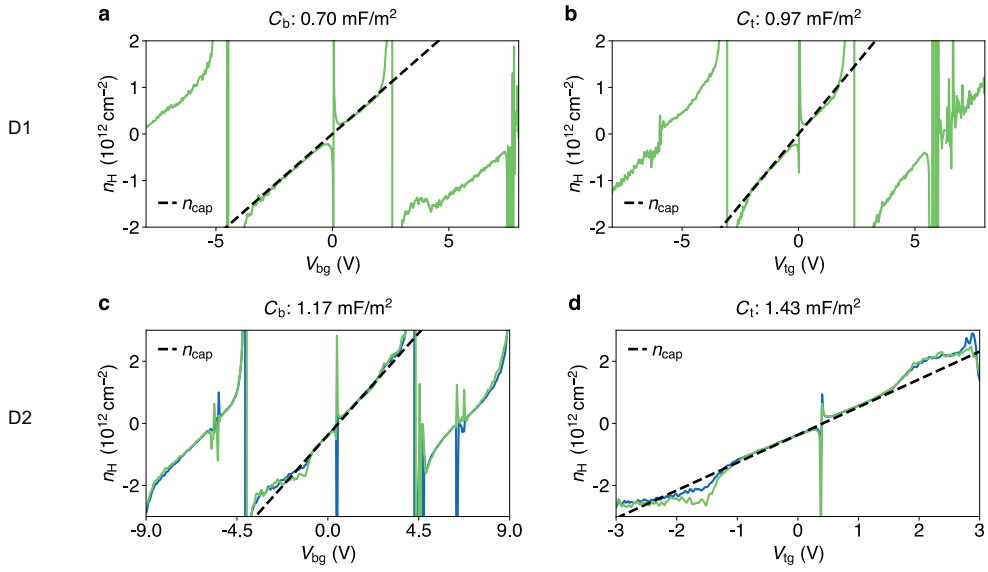

FIG. S6. **Gate capacitance calibration at** $T = 4$ **K. a, b** Shows $n_H(V_{bg}, V_{tg} = 0 \text{ V})$ and $n_H(V_{bg} = 0 \text{ V}, V_{tg})$ for device D1. **c, d,** Same as **a, b,** for device D2. The $n_H$ trace in green (blue) was antisymmetrized using $B_\perp = \pm 0.5$ T ($B_\perp = \pm 0.8$ T). $T = 4$ K in all of **a-d**.

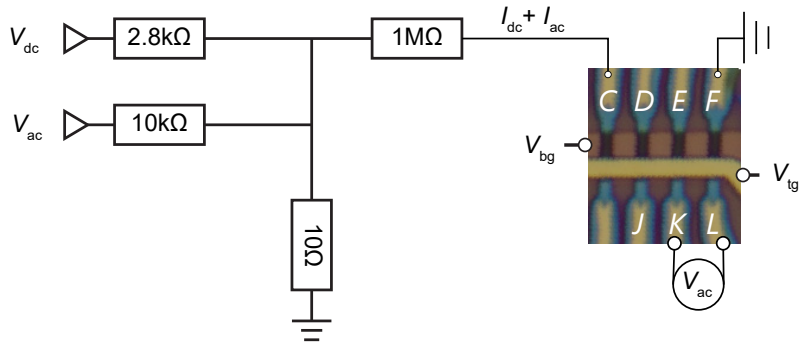

FIG. S7. **Circuit used to supply a AC + DC current bias.**

## C. Characterization of superconductivity

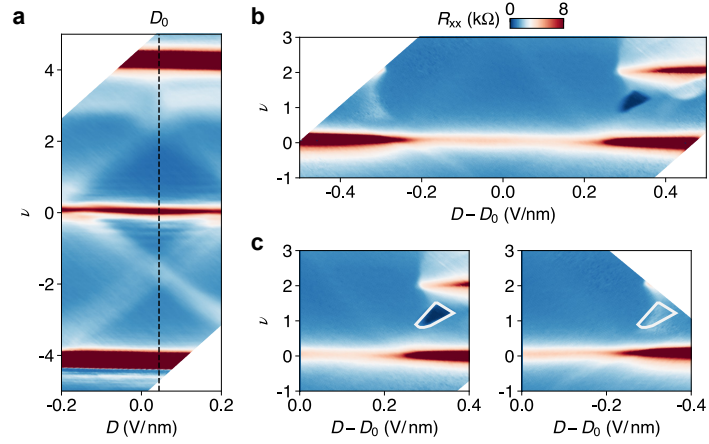

FIG. S8. **Absence of superconductivity for $D < 0$ in sample D1.** We show that a superconducting pocket is present only for $D > 0$ in device D1, given the range of gate voltage that can be applied prior to dielectric leakage. **a,** $R_{xx}$ as a function of applied electric displacement field, $D$, and band filling, $\nu$, obtained in $B_\perp = 0.8$ T. Cross-like resistive features were used to identify a small $D_0 = 0.044$ V/nm about which $R_{xx}(\nu, D)$ is qualitatively reflection symmetric. **b,** $R_{xx}(\nu, D - D_0)$ obtained at zero magnetic fields. **c,** A superconducting pocket (outlined in white) appears for $D - D_0$ between 0.285 to 0.360 V/nm and is absent when the sign of $D - D_0$ is reversed.

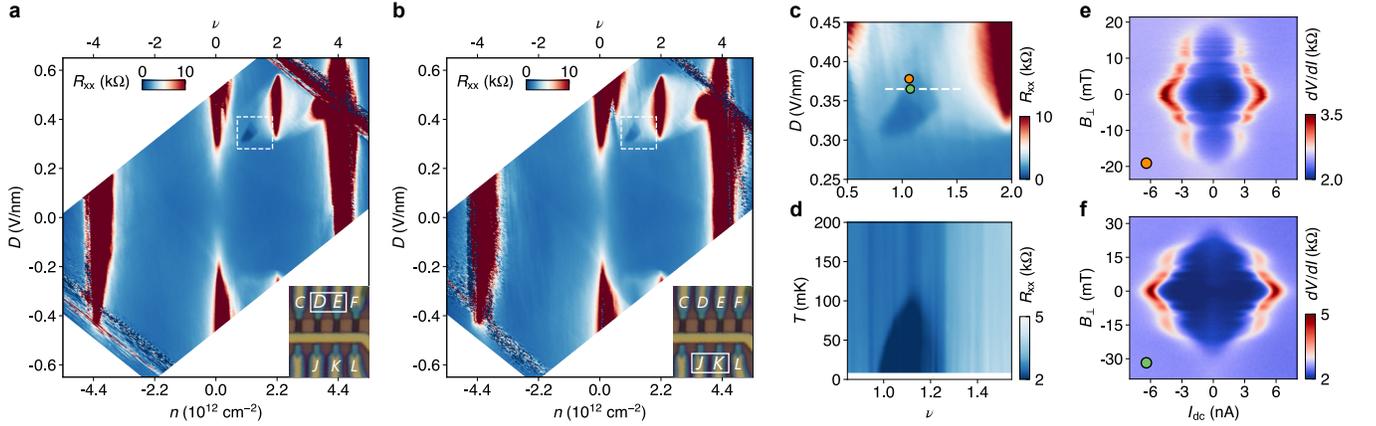

FIG. S9. **Signatures of superconductivity in other contact pairs in sample D1. a,b,** $R_{xx}(\nu, D)$ for two voltage probes D-E and J-K, with their locations shown in the inset, which also showed signatures of superconductivity (in addition voltage probe K-L, presented in the main paper). The low resistance pockets (resistance saturating to 0.45 and 1 kΩ respectively) are highlighted by a white dashed box. **c,** $R_{xx}(\nu, D)$ obtained inside the dashed box shown in **b**. **d,** Temperature dependence of the low resistance state taken along the dashed line in **b**, with $D = 0.365$ V/nm. **e, f,** Evolution of the differential resistance $dV/dI$ as a function of DC bias, $I_{dc}$, and perpendicular field, $B_\perp$. The oscillatory pattern of the $dV/dI$ is reminiscent of the Fraunhofer diffraction pattern, which suggests evidence of superconductivity as presented in the main manuscript. The corresponding location $\nu, D$ are marked by the orange and green markers in **c**, corresponding to $\{1.06, 0.378$ V/nm$\}$ and $\{1.08, 0.365$ V/nm$\}$ respectively.

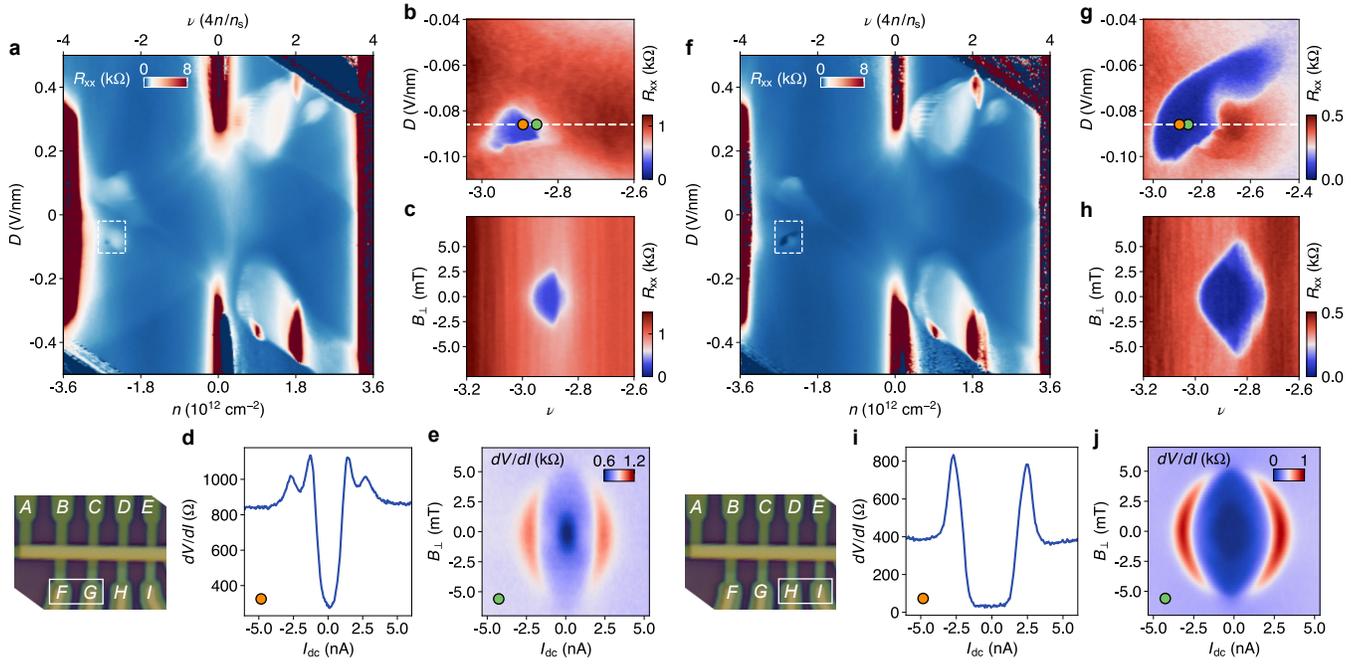

FIG. S10. **Signatures of superconductivity in sample D2.** Both voltage probe pairs F-G and H-I located on two ends of the hall bar showed signatures of superconductivity. **a,** $R_{xx}$ color plot showing the location of the superconducting pocket in $\nu$ and $D$. **b,** $R_{xx}(\nu, D)$, focusing on the superconducting region. **c,** A small $B_\perp$ rapidly suppresses the low resistance state, obtained along $D = -0.086$ V/nm. **d,** Differential resistance, $dV/dI$, as a function of $I_{dc}$, obtained at $\{-2.89, -0.086$ V/nm$\}$, indicated by the orange marker indicated in **b**. **e,** Differential resistance as a function of $I_{dc}$ and $B_\perp$ at $\{\nu, D\} = \{-2.86, -0.086$ V/nm$\}$, indicated by the green marker in **b**. **f-j,** Analogous to **a-e**, obtained using H-I as the voltage probes. Panel **f** is the same as Fig. 1c and **j** is the same as Fig. 2f of the main manuscript.

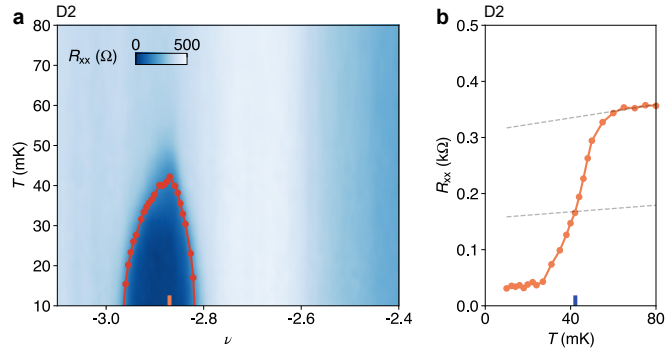

FIG. S11. **Temperature dependence of superconductivity in D2. a,** $R_{xx}(\nu, D = 0. - 0.08 \text{ V/nm})$ obtained at different temperatures. The red markers indicate $T_c$ extracted using a 50% resistance threshold. The highest $T_c$ indicated by the red markers was $\sim 42$ mK at $\nu = -2.87$. **b,** Line cut of $R_{xx}(T)$ at $\nu = -2.86$ (orange marker in **a**). The dashed lines indicate a linear fit to the high-temperature portion of the data to $r(T) = AT + B$. The intersection of $R_{xx}(T)$ with $r(T)/2$ was used to determine the $T_c$.

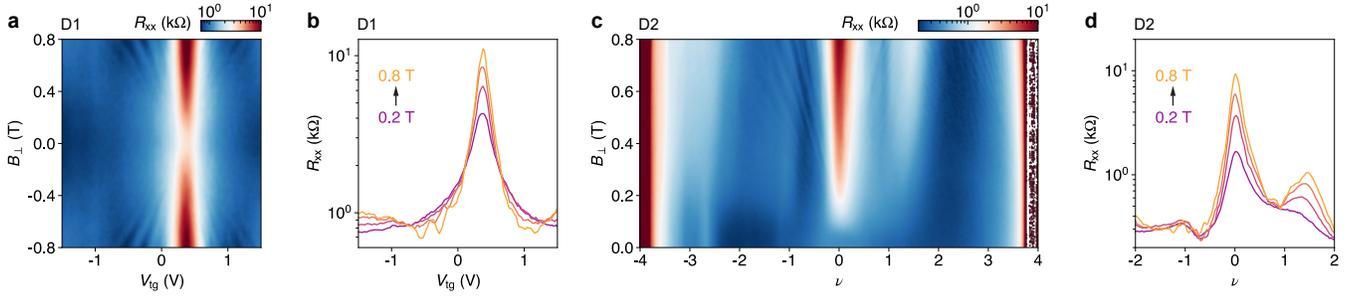

FIG. S12. **Shubnikov–de Haas oscillations. a,** The low field quantum oscillations in $R_{xx}$ as a function of top gate voltage and $B_\perp$ in D1 around $D = 0$ V/nm. We observe sDH oscillations appearing at $B_\perp \sim 400$ mT. **b,** Shows line traces for several values of in $B_\perp$ in **a**. **c,** Similar measurements in D2, taken as a function of band filling for $D = -0.08$ V/nm. We observe sDH oscillations appearing at $B_\perp \sim 200$mT. **d,** Line cuts from **c**.

## D. Coherence Length Analysis

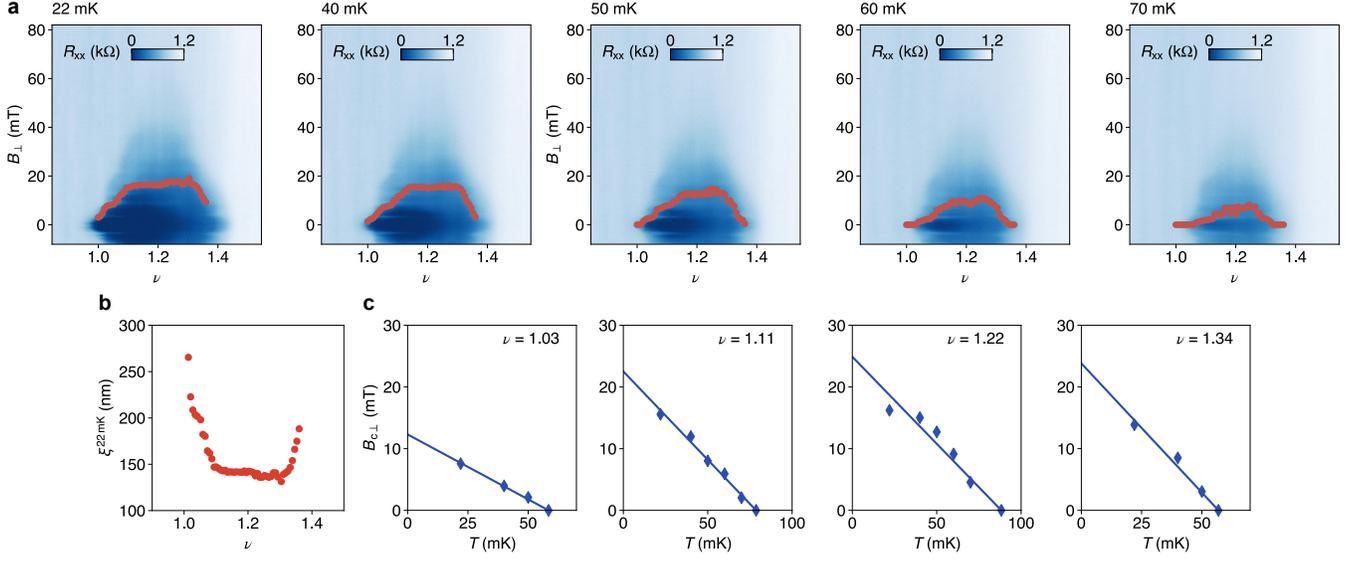

FIG. S13. **Band filling dependent coherence length analysis for sample D1. a,** $R_{xx}(\nu, D = 0.365 \text{ V/nm})$ as a function of an out-of-plane magnetic field, $B_\perp$, for several temperatures. The red markers indicate the out-of-plane critical field, $B_{c\perp}$, estimated by fitting high field part of $R_{xx}(B_\perp)$ at each $\nu$ to $r(B_\perp) = AB_\perp + C$, and finding the intersection of $R_{xx}(B_\perp)$ with $r(B_\perp)/2$. **b,** Ginzburg Landau coherence length, $\xi$, estimated using $B_{c\perp}$ at 22 mK as a function of $\nu$, according to $\xi = \sqrt{\Phi_0/(2\pi B_{c\perp})}$, where $\Phi_0 = h/2e$ is the magnetic flux quantum. The estimated $\xi^{22 \text{ mK}} \sim 150$ nm is consistent with $\xi_{GL}$ estimated using the extrapolated out-of-plane critical field at $T = 0$, $B_{c\perp}^0$, shown in Fig. 3b of the main manuscript. **c,** Temperature dependence of $B_{c\perp}$ for representative values of $\nu$ across the superconducting dome, showing the estimation of $B_{c\perp}^0$ by fitting to the Ginzburg Landau relation $T/T_{c,0} = 1 - (2\pi\xi_{GL}^2/\Phi_0)B_{c\perp}$.

## E. In-Plane Magnetic Field Dependence and Pauli Limit Analysis

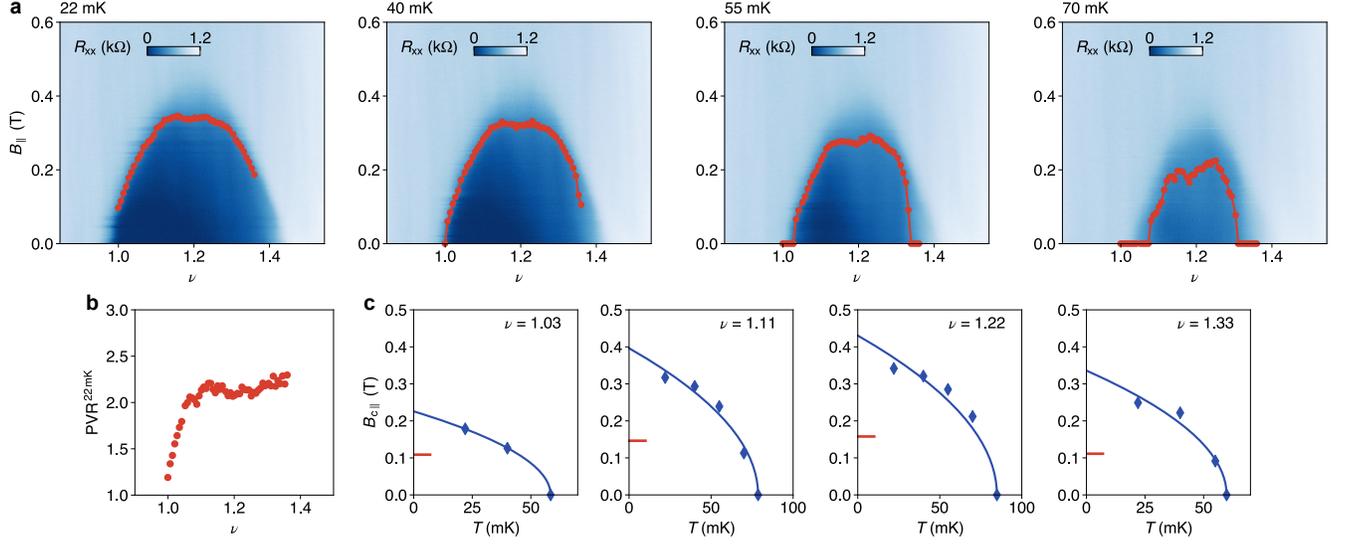

FIG. S14. **Band filling dependent Pauli limit violation analysis for sample D1. a,** $R_{xx}(\nu, D = 0.365$ V/nm) as a function of an in-plane magnetic field, for several temperatures. The red markers indicate the in-plane critical magnetic field, $B_{c\parallel}$, by fitting the high field part of $R_{xx}(B_\parallel)$ to $r(B_\parallel) = AB_\parallel + C$, and finding the intersection of $R_{xx}(B_\parallel)$ with $r(B_\parallel)/2$. **b,** Pauli limit violation ratio, $B_{c\parallel}/B_p$, estimated using $B_{c\parallel}(\nu)$ found at $T = 22$ mK, and $B_p \sim 1.86$ T/K $\times T_c(\nu)$, where $T_c$ was defined using a 50% resistance threshold. PVR$^{22\text{ mK}}$ is consistent with PVR between 2 to 3 across most of the superconducting dome shown in Fig.3c of the main manuscript. **c,** Extrapolation of $B_{c\parallel}^0$ using the phenomenological relation $T/T_c^0 = 1 - (B_{c\parallel}/B_{c\parallel}^0)^2$. The red markers along the $B_{c\parallel}$ axis show the estimated Pauli limit field at the particular band filling, $D = 0.365$ V/nm.

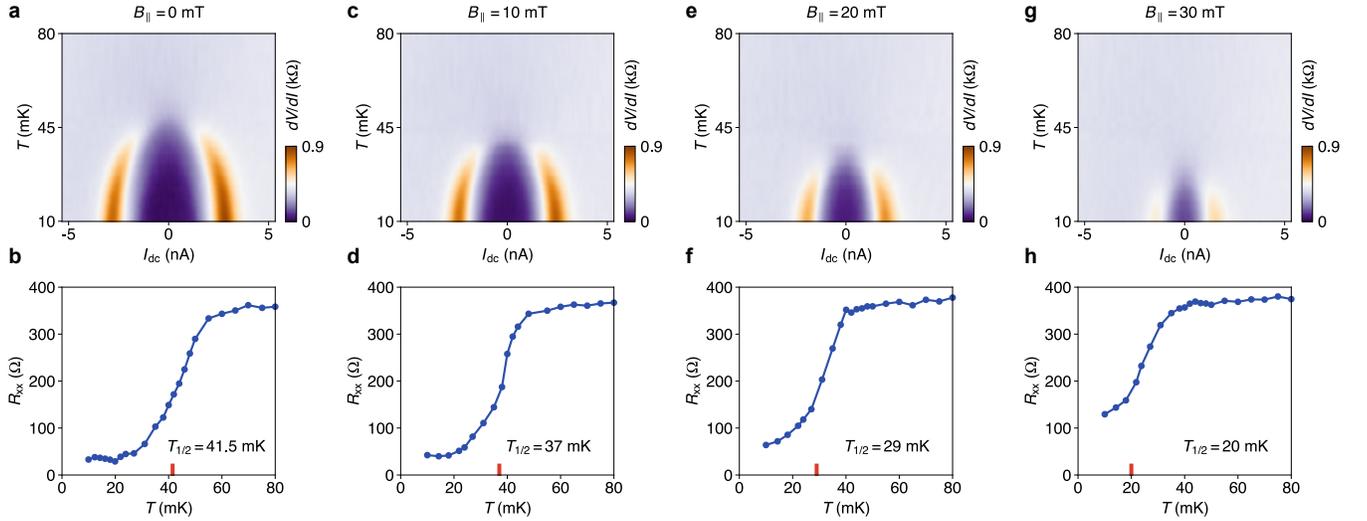

FIG. S15. **In-plane magnetic field and temperature dependence in D2. a, b,** Temperature dependence of non-linear $dV/dI$. A line cut at $I_{dc} = 0$ was used to find $T_c$ defined by the 50% normal state resistance threshold. Panels (**c, d**), (**e, f**), and (**g, h**) are analogous to **a, b,** but for increasing applied in-plane magnetic field. Obtained at $\nu = -2.86$, $D = -0.08$ V/nm, at the same location as Fig. 3f of the main manuscript.

## F. Magnetic Field Dependence of Correlated Insulating States

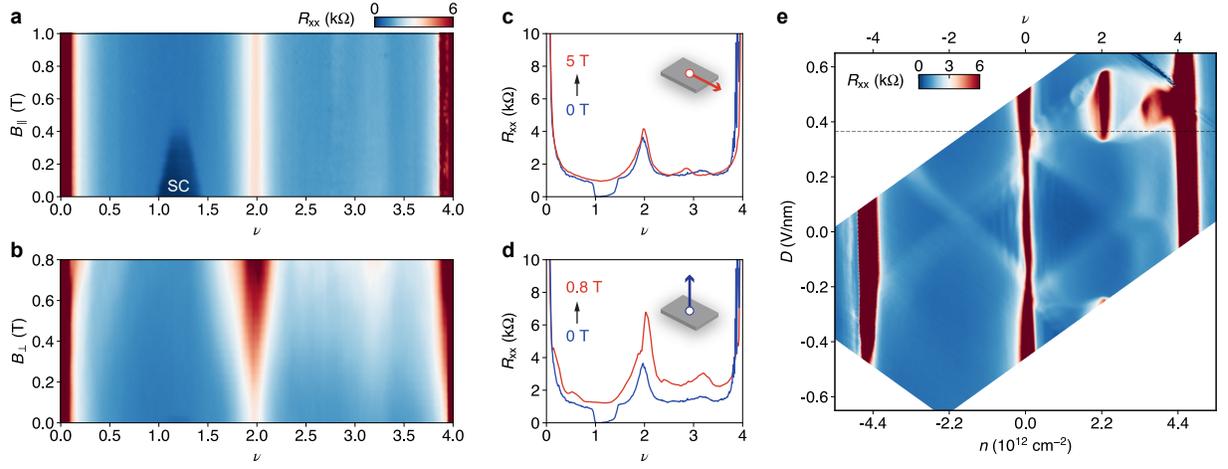

FIG. S16. **Magnetic field response of correlated insulating states in D1. a,** Color map of $R_{xx}(\nu, D = 0.365$ V/nm) as a function of in-plane magnetic field, $B_{\parallel}$. **b,** $R_{xx}(\nu, D = 0.365$ V/nm) as a function of out-of-plane magnetic field, $B_{\perp}$. **c,** Comparison between $R_{xx}$ acquired at $B = 0$ and $B_{\parallel} = 5$ T. **d,** Comparison between $R_{xx}$ at $B = 0$ and $R_{xx}$ field symmetrized at $B_{\perp} = \pm 0.8$ T. **e,** Field symmetrized $R_{xx}(\nu, D)$ obtained in $B_{\perp} = \pm 0.8$ T. A resistance peak develops near $\nu = 3$ at $D \sim 0.44$ V/nm.

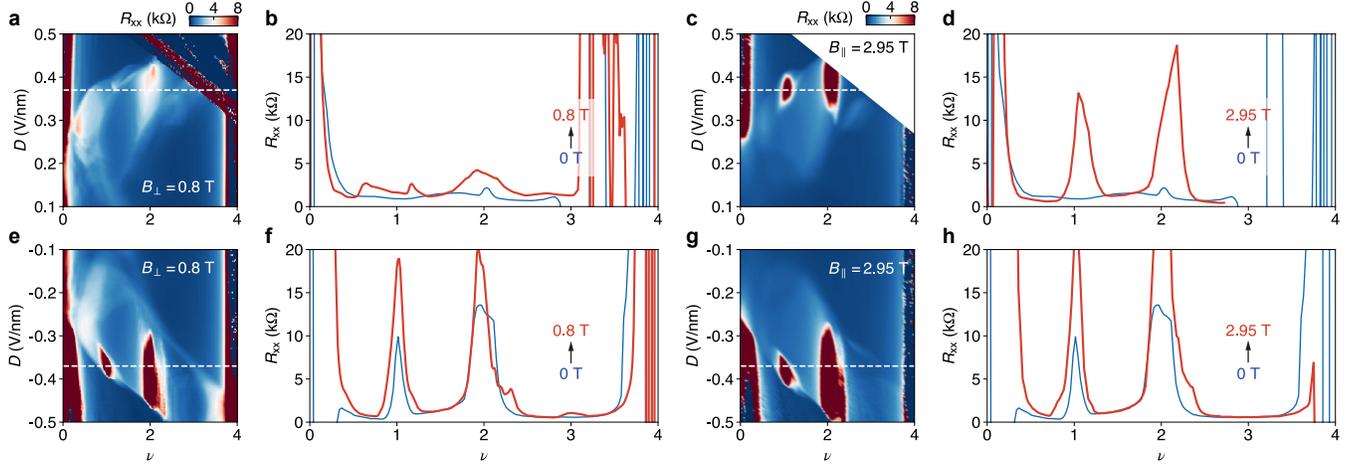

FIG. S17. **Magnetic field response of correlated insulating states in D2. a,** Field symmetrized $R_{xx}(\nu, D)$ obtained at $B_{\perp} = \pm 0.8$ T, focusing on the $D > 0$ region near the $\nu = 2$ insulating of device D2. **b,** Comparison between line cuts of $R_{xx}(\nu, D = 0.37$ V/nm) at $B = 0$ and $B_{\perp} = 0.8$ T along the white dashed line in **a**. **c,** $R_{xx}(\nu, D)$ obtained in $B_{\parallel} \sim 3$ T, focusing on the same region as **a**. **d,** Comparison between $R_{xx}(\nu, D = 0.37$ V/nm) at $B = 0$ and $B_{\parallel} \sim 3$ T. **e,** $R_{xx}$ obtained under the same condition as **a**, focusing on $D < 0$. **f,** Comparison between $R_{xx}$ line cuts made at $D = -0.37$ V/nm. **g, h,** Analogous data to **c, d**, focusing on the insulating state at $D < 0$. In contrast to sample D1 (Fig. S16), an in-plane magnetic field has a much greater effect on the resistance of $\nu = 2$ insulating state, consistent with a spin-polarized ground state.

## G. Signatures of field-induced isospin symmetry breaking in D2

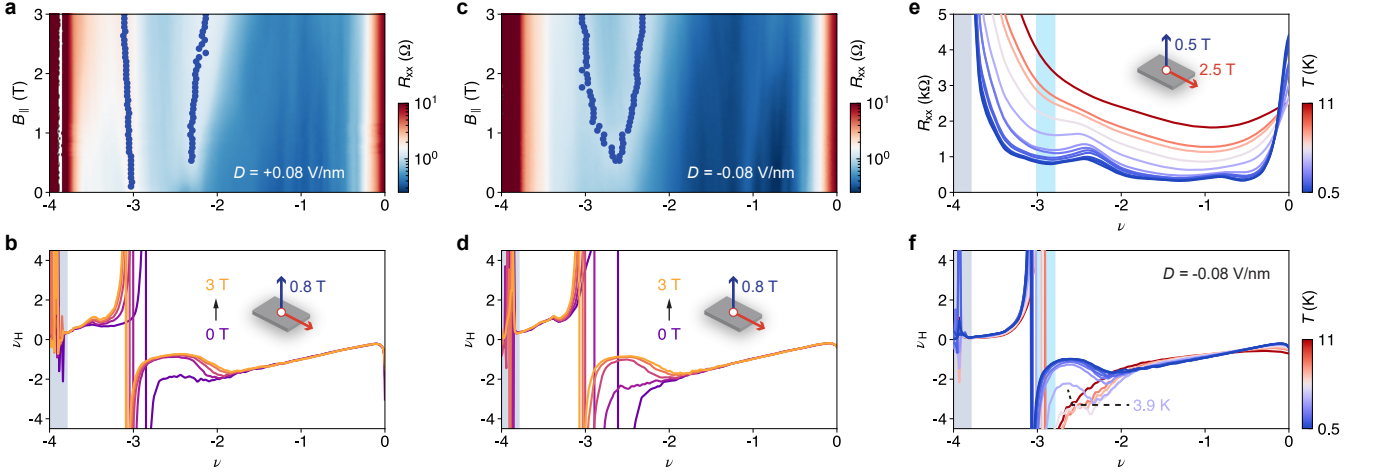

FIG. S18. **Signatures of field-induced isospin symmetry breaking in D2. a, b,** and **c, d,** contrasts the behavior of $R_{xx}(\nu)$ and $\nu_H(\nu)$ for +0.08 V/nm and −0.08 V/nm in $B_\perp = 0.8$ T as a function of $B_\parallel$ from 0 to 3 T respectively. In **a, c,** two resistive features (peak positions highlighted by blue markers) encloses isospin polarized region for $\nu$ between −3 to −2, where the Hall filling factor undergoes a reset-like behavior. While the resistive features in **a** extend down to $B_\parallel = 0$, the resistive features in **c** develop only when $B_\parallel > \sim 1$ T is applied. Similarly, while signatures of a $\nu_H$ reset is present for $D = +0.08$ V/nm, even at $B_\parallel = 0$, a reset feature at $D = -0.08$ V/nm appears only when a finite $B_\parallel$ was applied. Combined with the observation that superconductivity is absent at +0.08 V/nm, but present at -0.08 V/nm, our findings suggest that the superconducting phase observed is enhanced when the system is on the cusp of isospin symmetry breaking but suppressed when the symmetry is fully broken. **e, f,** temperature dependence of $R_{xx}$ and $\nu_H$ along constant $D = -0.08$ V/nm, obtained in $B_\perp = 0.5$ T and $B_\parallel = 2.5$ T. **b, d** are line cuts from Fig. 4i of the main manuscript.

## H. Additional low field hall density measurements

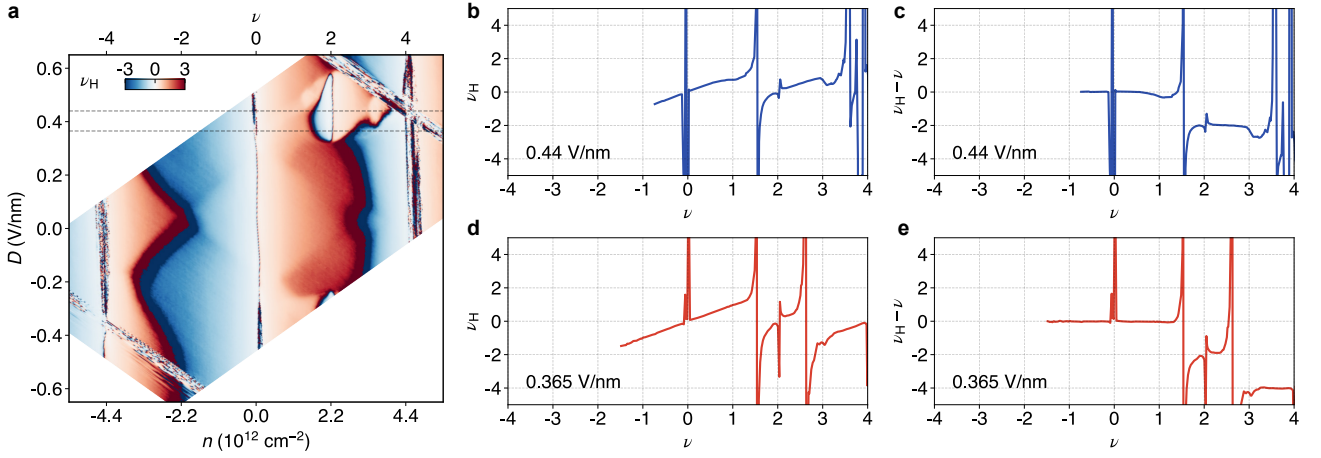

FIG. S19. **Hall filling measurements for device D1. a,** Antisymmetrized hall filling obtained at $B_\perp = \pm 0.8$ T (same as Fig. 4a of main manuscript). **b, c** and **d, e** are line cuts of $\nu_H$ and $\nu_H - \nu$ for two values of $D$ indicated by the black dashed lines in **a**.

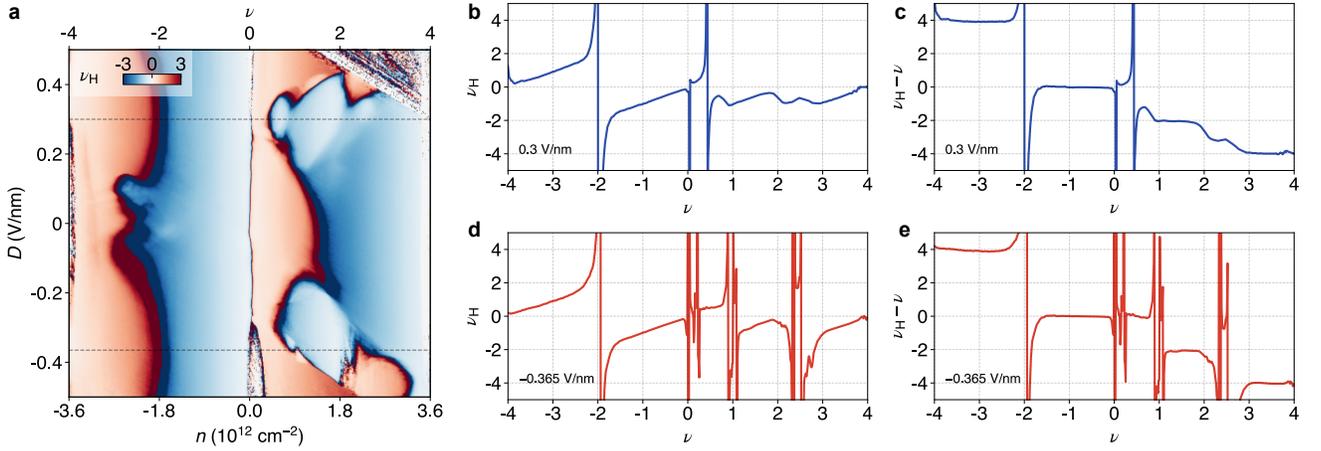

FIG. S20. **Hall filling measurements for device D2. a,** Antisymmetrized hall filling obtained at $B_\perp = \pm 0.8$ T (same as Fig. 4f of main manuscript). For $D$ near $\sim \pm 0.2$ V/nm, starting at $\nu = 0$, the hall filling first passes through a van Hove singularity near $\nu = 1$, then undergoes several resets near integer $\nu$, until $\nu = 4$. As $D$ is increased to $\sim 0.4$ V/nm, the reset at $\nu = 2$ is replaced by a change in sign of $\nu_H$ where a correlated gap opens. **b, c** and **d, e** are line cuts of $\nu_H$ and $\nu_H - \nu$ along the black dashed lines in **a**.

## I. Example of Weak Localization in D2

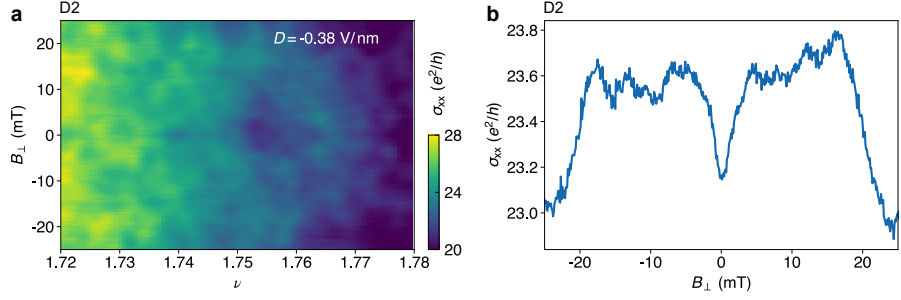

FIG. S21. **Weak localization in D2.** This figure illustrates an example of what we believe to be a weak localization correction in D2, in a halo region in the conduction band. Similar data was collected across the gate-tuned phase diagram, although in many regions no zero-field correction attributable to weak localization was visible. We found no clear pattern in where the correction was or was not visible. Never was weak anti-localization (a peak in conductance at B=0) observed. This data was taken at mixing chamber temperature 10mK. **a,** The color plots of four terminal conductivity $\sigma_{xx}$ in units of $e^2/h$, as a function of $B_\perp$ and $\nu$, obtained at $D = -0.38$ V/nm, for the same voltage probes (H-I) used for measurements of superconductivity. **b,** Averaged $\sigma_{xx}$ over the range of filling factors in **a**. The suppression of $\sigma_{xx}$ by $\sim e^2/2h$ at $B_\perp = 0$ is attributed to weak localization.

# SUPPLEMENTARY REFERENCES

23



\end{document}